# A Novel Bluetooth Man-In-The-Middle Attack Based On SSP using OOB Association model

[1]K.Saravanan, [2]L.Vijayanand and [3]R.K.Negesh,
[1]Lecturer/CSE and [2,3]Lecturer/EEE
[1,2]Erode Sengunthar Engineering college and [3]Marthandam college of Engineering and Technology,
[1]saravanankumarasamy@gmail.com [2]vijay.21a@gmail.com and [3]rknegesh@yahoo.com

**ABSTRACT**

*As an interconnection technology, Bluetooth has to address all traditional security problems, well known from the distributed networks. Moreover, as Bluetooth networks are formed by the radio links, there are also additional security aspects whose impact is yet not well understood. In this paper, we propose a novel Man-In-The-Middle (MITM) attack against Bluetooth enabled mobile phone that support Simple Secure Pairing( SSP). From the literature it was proved that the SSP association models such as Numeric comparison, Just works and passkey Entry are not more secure. Here we propose the Out Of Band (OOB) channeling with enhanced security than the previous methods.*

## I. INTRODUCTION

Bluetooth is an open standard for short-range radio frequency (RF) communication. Bluetooth standard specifies wireless operation in the 2.4 GHz–2.4835 GHz ISM frequency band and supports data rates up to 720 Kbps. Bluetooth uses a frequency-hopping spread-spectrum (FHSS) technology to solve interference problems. The FHSS scheme uses 79 different radio channels by changing frequency about 1,600 times per second. This allows users to form ad-hoc networks between a wide variety of devices to transfer voice and data. Bluetooth eliminate wires and cables between both stationary and mobile devices. Bluetooth is a low-cost, low-power technology that provides a mechanism for creating small wireless networks on an ad-hoc basis, known as piconets. A piconet is composed of two or more Bluetooth devices in close physical proximity that operate on the same channel using the same frequency hopping sequence.

Because Bluetooth is a wireless communication system, there is always a possibility that its transmissions could be deliberately jammed or intercepted, or false/altered information could be passed to the piconet devices. To provide protection for the piconet, the system can establish security at several protocol levels. Bluetooth has built-in security measures at the link level.

Our work mainly concentrates on the Man-In-The-Middle attack. By principle, without any verification of the public keys, MITM attacks are generally possible against any message sent by using public-key technology. The existing model uses the Bluetooth device that support SSP (Printer) that makes use of the Just Works, Numeric Comparison and the Pass key entry association models. But it was proved that the existing model is not very much secure [1]. So we propose to use Out-Of-Band channeling association model to have more security.

Out Of Band refers to communications which occur outside of a previously established communication methods or channel. The cryptographic systems that are secure against MITM attacks require an additional exchange or transmission of information over some kind of secure channel.

The paper is organized as follows. Section II provides an overview of Bluetooth security. Section III gives how secure simple pairing works and its authentication stages. The existing method and its possibility to MITM attack is provided in Section IV. Our work is proposed in Section V. Section VI gives the results and discussions we have gone with so far and about the results we expect. Conclusion is given in section VII.

# II. BLUETOOTH SECURITY

Principles of security include
- Authentication- Identity verification
- Data Confidentiality- protection of transmitted data from passive attacks.
- Authorization-allow the control of resources.

Bluetooth security configuration is done by the user who decides how a Bluetooth device will implement its connectability and discoverability options. The different combinations of connectability and discoverability capabilities can be divided into three modes. Each Bluetooth device can operate in one mode only at a particular time. The three modes are the following:

Security Mode 1—Non-secure mode. , In this mode, the security functionality (authentication and encryption) is completely bypassed. It just monitors traffic.
Security Mode 2—Service-level enforced security mode, security procedures are initiated after channel establishment at the Logical Link Control and Adaptation Protocol (L2CAP) level.
Security Mode 3—Link-level enforced security mode, a Bluetooth device initiates security procedures before the channel is established. This is a built-in security mechanism. This mode supports authentication and encryption. These features are based on a secret link key that is shared by a pair of devices. To generate this key, a pairing procedure is used

Bluetooth related threats include Bluesnarfing, Blue jacking, and Blue bugging etc. An example of a long-distance attacking tool is the Blue Sniper Rifle. It is a rifle stock with a powerful directional antenna attached to a small Bluetooth compatible computer. At one side the security measures for Bluetooth devices increase, but on the other side new methods to crack the security measures are discovered.

In order for two Bluetooth devices to start communicating, procedure called *pairing* must be performed. As a result of pairing, two devices form a trusted pair and establish a link key which is used later on for creating a data encryption key for each session. Simple Pairing has two security goals: protection against passive eavesdropping and protection against MITM attacks (active eavesdropping). It is also a goal of Simple Pairing to exceed the maximum security level provided by the use of a 16 character alphanumeric PIN with the pairing algorithm. In Bluetooth versions up to 2.0+EDR, pairing is based exclusively on the fact that both devices share the same *PIN (Personal Identification Number)* code or passkey. When the user enters the same passkey in both devices, the devices generate the same shared secret which is used for authentication and encryption of traffic exchanged by them. Even with longer 16-character alphanumeric PINs, full protection against active eavesdropping cannot be achieved: it has been shown that MITM attacks on Bluetooth communications (versions up to 2.0+EDR) can be performed [3], [4].

Bluetooth version 2.1+EDR adds a new specification for the pairing procedure, namely Secure Simple Pairing (SSP). Its main goal is to improve the security of pairing by providing protection against passive eavesdropping and MITM attacks. SSP employs Elliptic Curve Diffie-Hellman public-key cryptography. To construct the link key, devices use public-private key pairs, a number of nonces, and Bluetooth addresses of the devices. 2.1+ EDR versions have got an additional mode along with the given three modes. This security mode 4 is defined for Secure Simple Pairing.

SSP works on association models which can be defined according to the I/O capabilities of the two devices that are connected [5].

• Numeric Comparison: designed to be used when both devices are capable of displaying a six-digit number and user input of "yes" or "no". A typical example could be two cell phones pairing with each other. The six-digit number displayed in this model is an output of the underlying security algorithm.
• Just Works: designed to be used when at least one of the devices does not have display capability of six digits and also is not capable of entering six decimal digits using a keyboard or any other means. This model does not provide protection against MITM attacks. Compared to the legacy headsets with a fixed PIN, the security level provided by this model is much higher.
• Out of Band: designed for devices capable of using Out of Band (OOB) mechanisms to exchange secretes to be used in the pairing process. Near Field Communication (NFC) to exchange the cryptographic information by touching two devices is an example.
• Passkey Entry: designed to be used when one of the devices does not have display capability of six digits, but has the input capability and the other device has output capability. PC and keyboard is a typical example of this model. The user is shown a six-digit number and then asked to enter from the device with only input capability.

# III. SECURE SIMPLE PAIRING STAGES

The authentication stages based on SSP have four main stages. Prior to the stages first the association model is decided according to the I/O capabilities of the two devices that getting interacted. The next step here is the private key and the public key generation. This step also helps in computing the Diffie-Hellman key. The four main stages of authentication using SSP are as follows [6], [7].

• Connection establishment: The user on the SAP Client end will select one of the device possible SAP Server to connect with and perform Bluetooth Service Discovery to be absolutely

•Authentication Stage 1: This stage varies slightly for numeric comparison, OOB and passkey entry. Let us assume, that based on the IO Capability of both the devices, one side displays a six digit number for the other device to enter. After this stage the Authentication Stage 2 starts. Figure 1 explains this concept with OOB. Table I explains about the notations used.

• Authentication Stage 2: At this stage, the results of the cryptographic functions are compared and if it is successful, the host receives Simple Pair Complete event and then the link key computation starts.

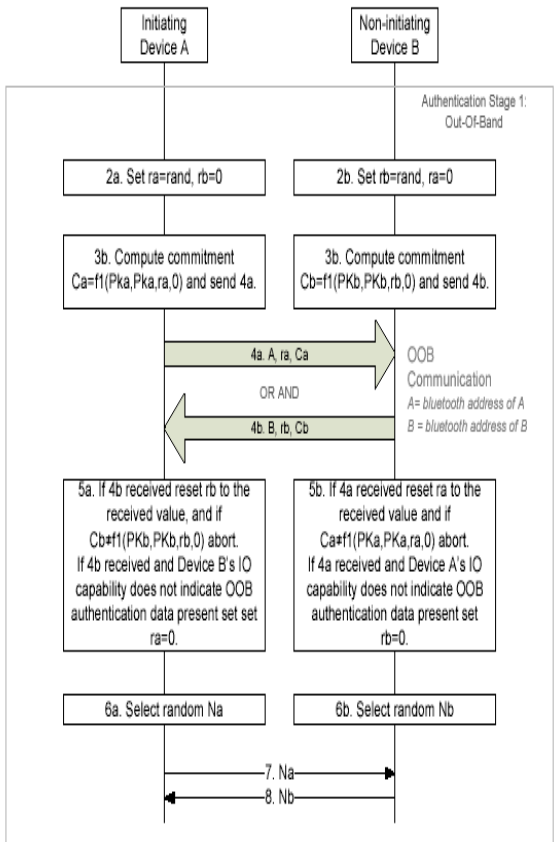

Figure 1. Authentication stage 1 with OOB

TABLE I
PROTOCOL NOTATION

| Term | Definition |
|---|---|
| PKx | Public key of device X |
| Nx | Nonce generated by device X |
| rx | Random number generated by device X |
| Cx | Commitment value from device X |
| f1 | One-way function used to compute commitment values |
| IO | Input/Output capabilities |

• Link key calculation: At this stage, a Link Key is calculated and mutual authentication is performed to make sure both the devices have the same link key. Similar to the legacy implementation, Link Key Notification event is generated on both the initiator and responder of the Simple pairing procedure. On the initiator end, Authentication Complete event is generated at the end of this phase.

• Enable Encryption: The encryption phase is the same as the legacy implementation. The link level encryption can be enabled using HCI_Set_Connection_Encryption command. Once the link is encrypted, the SAP Client will start the L2CAP Channel establishment procedure for RFCOMM. Finally, SAP connection will be made over the RFCOMM Channel.

Even though version 2.1+EDR provides greater security measures against the attacks, the devices that are commonly used in practice is version 2.0+EDR and it is proved that MITM attack is possible in this version.

## IV. EXISTING METHOD

Consider a Bluetooth enabled device that support SSP for example a Bluetooth enabled mobile phone here. The association model used here is Just Works. The MITM attack in the Bluetooth enabled devices is explained with the help of the figure 2 [2], [8].

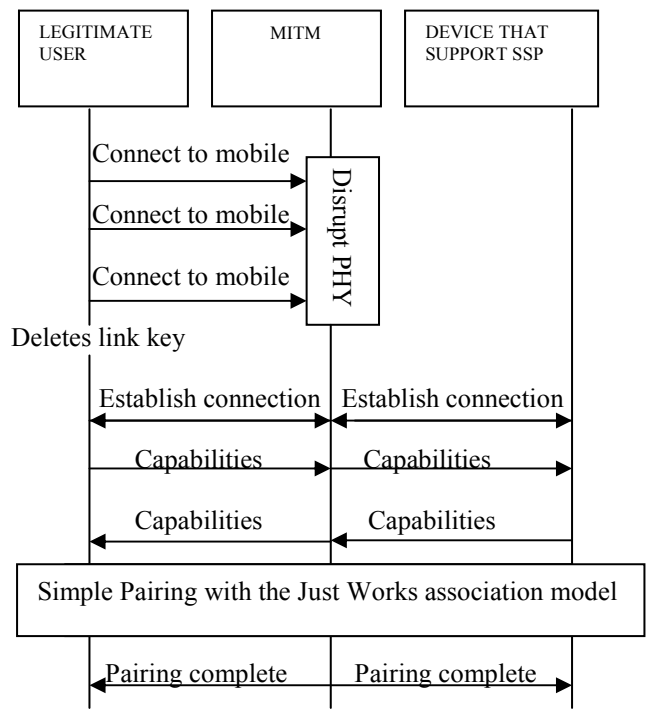

Figure 2. Idea of the attack

The attack is explained in two scenarios. In the first scenario the victim devices use the Just Works association model. Our attack works in the following way:

1) The MITM jams the physical layer by hopping along with the victim devices and sending random data in every timeslot
2) The MITM impersonates the legitimate device.
3) MITM intercepts all data.
4) The MITM relays the data to the legitimate device.

The second scenario is that the devices use the Numeric Comparison or the Pass-key Entry model. Both the methods give a conclusion that the MITM attack is possible. So we propose to use the OOB association model and to use Near Field Communication as the OOB channel.

The link key derived here is well protected since it is derived from a Diffie-Hellman key exchange (carried out in an Elliptic curve group, for greater efficiency). However, a huge vulnerability is found in the pairing mode that is based on passkey entry, numeric comparison, or the just works model.

## V. PROPOSED METHOD

The security against attacks like MITM attack is to be improved. Near Field Communication as the Out Of Band channeling is the best suggested method. Just works association model is used as an optional one and the Out Of Band as the mandatory model.

Such devices that cannot use the new window at the user interface level or alternatively NFC as an OOB channel (better way), should implement their security either in the same way as old Bluetooth devices (versions up to 2.0+EDR) do or not to use Bluetooth security at all (if no sensitive data is exchanged). In this way, the implementation of the Just Works association model can be made optional and perhaps even removed altogether from the Bluetooth SSP specification. The advantage of this approach is to eliminate all MITM attacks against the Just Works association model. Moreover, if the Just Works association model is not supported anymore in the future Bluetooth devices, it is not possible to force victim devices to use it.

Future Bluetooth specifications should make OOB a mandatory association model in order to radically improve the security and usability of SSP. However, it is likely that such a radical change in the specification will not be possible at once. Therefore, future Bluetooth specifications should at least strongly recommend the use of an OOB channel (e.g., Near Field Communication) to all Bluetooth device manufacturers.

Use of the private security level, increasing user awareness of security issues, minimization of transmit powers, careful selection of place where sensitive information is exchanged are some other efficient methods that can be encouraged to improve the security of SSP. Among these methods the OOB channeling is the one that is proposed in this paper.

To provide more security against MITM attack with the OOB channeling various frequencies are considered so that the Bluetooth device while communicating uses not the same OOB frequency all the time of its communication. It uses different frequencies at each time of its communication.

## VI. RESULT AND DISCUSSION

The secure simple pairing process is done using JAVA programming. First the pairing is done from PC to PC which is shown in the figure 3 that is the encryption and decryption.

Our further work is on the improvement to the Secure Simple Pairing. Out Of Band association model is to be developed and proved it to be an efficient method.

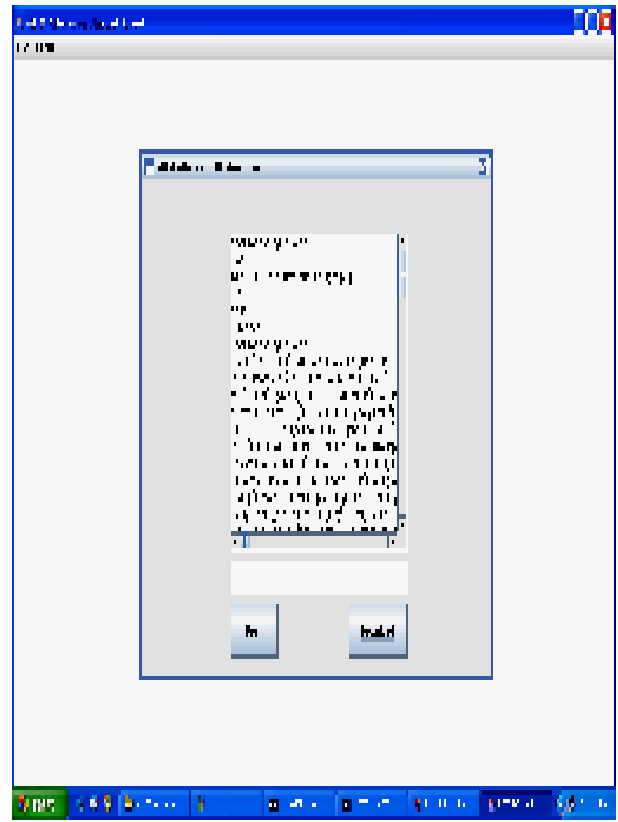

Figure 3. PC to PC pairing

## VII. CONCLUSION

It is difficult to create a protocol which caters to all possible types of wireless devices, as the security of the protocol is likely to be limited by the capabilities of the least powerful or the least secure device type. Our Bluetooth MITM attack presented in this paper is based on this problem. By far the best way to stop the attacks is to use an OOB channel, and SSP supports this option. Near Field Communication is used as the OOB channel and going to prove that this method is more secure.

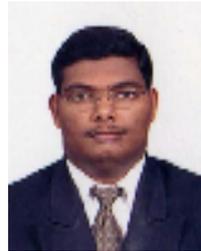
**Mr.K.Saravanan** received the M.E degree 2008 in computer science from Dr.MCET, Anna University, Chennai, India. He is currently working as a Lecturer at the Faculty of Engineering, Erode Sengunthar Engineering College, Erode, Tamilnadu. He has published 4 paper in International Journal, 07 papers in National Conference and 02 papers in International Conference..His current research interests are information security, computer communications and DDoS Attacks. He is currently pursuing Ph.D. under Anna University of Technology, Coimbatore.

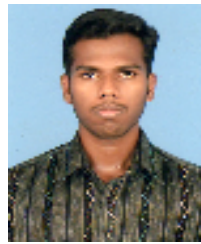
**L. Vijay Anand** received the BE Degree in Electrical and electronics from Sun Engineering college, Erachakulam, Nagercoil in 2007 and the ME Degree in Applied Electronics from the Bannari Amman Institute of Technology, Sathyamangalam in 2009. He is Lecturer, working in Erode Sengunthar Engineering College, Erode. His current research focuses on mobile ad hoc networks and wireless security.

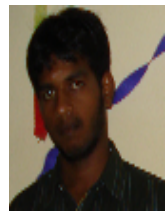
**R.K. Negesh** received the BE Degree in Electrical and electronics from Sun Engineering college, Erachakulam, Nagercoil in 2007 and the ME Degree in Applied Electronics from the Bannari Amman Institute of Technology, Sathyamangalam in 2009. He is Lecturer, working in Marthandam college of Engineering and Technology. His current research focuses on security of Adhoc networks